\documentclass[final,authoryear,5p,times,twocolumn]{elsarticle-arxiv}

\usepackage[utf8]{inputenc}
\usepackage[T1]{fontenc}
\usepackage[english,francais]{babel}

\usepackage{amssymb}

\usepackage[colorlinks,bookmarks=false]{hyperref}
\usepackage{url}

\usepackage{tikz}

\usetikzlibrary{intersections}

\usepgfmodule{oo}

\pgfooclass{treeplet}{
    \attribute scale;
    
    \method treeplet(#1){
    \pgfooset{scale}{#1}
    }
    
    \method draw(#1){
    \begin{scope}[scale=\pgfoovalueof{scale}]
        \node (T) at (#1) {};
        \path (T) -- ++(0,1) node (S) {};
        \path (T) -- ++(0,-1) node (B) {};
        \path (B) -- ++(-1,0) node (A) {};
        \path (B) -- ++(1,0) node (C) {};
        \draw[very thick] (A.center) -- (S.center) node[pos=0.5] (AB) {};
        \draw[very thick] (AB.center) -- (B.center);
        \draw[very thick] (S.center) -- (C.center);
    \end{scope}
    }
}

\pgfooclass{target}{
    \attribute scale;
    
    \method target(#1){
    \pgfooset{scale}{#1}
    }
    
    \method draw(#1){
    \begin{scope}[scale=\pgfoovalueof{scale}]
        \node (T) at (#1) {};
        \draw[thick] (T) circle [x radius=1.5cm,y radius=3cm];
        \draw[thick] (T) circle [x radius=1cm,y radius=2cm];
        \draw[thick] (T) circle [x radius=0.5cm,y radius=1cm];
    \end{scope}
    }
}

\usepackage{pgfplots}
\pgfplotsset{every axis legend/.append style={cells={anchor=west},at={(0.01,0.99)},anchor=north west}}

\colorlet{T}{blue}

\colorlet{C}{green}

\colorlet{A}{red}

\colorlet{G}{yellow}

\colorlet{K}{G!50!T}

\colorlet{S}{G!50!C}

\colorlet{Y}{T!50!C}

\colorlet{M}{A!50!C}

\colorlet{W}{A!50!T}

\colorlet{R}{G!50!A}

\colorlet{B}{G!33!T!33!C}

\colorlet{D}{G!33!A!33!T}

\colorlet{H}{A!33!C!33!T}

\colorlet{V}{G!33!C!33!A}

\colorlet{N}{A!25!G!25!C!25!T}

\colorlet{green}{black!50!green}
\colorlet{cyan}{black!25!cyan}
\colorlet{red}{black!25!red}
\colorlet{orange}{black!25!orange}

\def\soa{analytical strategy}
\def\soas{analytical strategies}

\def\ap{\emph{a posteriori}}
\def\gtrig{$\mathrm{GTR} {+} \mathrm{I} {+} \Gamma$}

\usepackage{comment}

\journal{Comptes Rendus Palevol}

\begin{document}

\begin{frontmatter}



\title{Evaluating strategies of phylogenetic analyses by the coherence of their results\\\large{\emph{Évaluation des stratégies d'analyse phylogénétique par la cohérence de leurs résultats}}}


\author[BL]{Blaise Li\corref{cor}}
\cortext[cor]{Corresponding author.}

\ead{blaise.li@normalesup.org}
\address[BL]{Centro de Ciências do Mar, Universidade do Algarve, Campus de Gambelas, 8005-139 Faro, Portugal}

\begin{abstract}

I propose an approach to identify, among several strategies of phylogenetic
analysis, those producing the most accurate results.  This approach is based on
the hypothesis that the more a result is reproduced from independent data, the
more it reflects the historical signal common to the analysed data.  Under this
hypothesis, the capacity of an \soa{} to extract historical signal should
correlate positively with the coherence of the obtained results.  I apply this
approach to a series of analyses on empirical data, basing the coherence measure on
the Robinson-Foulds distances between the obtained trees.  At first
approximation, the \soas{} most suitable for the data produce
the most coherent results.  However, risks of false positives and false
negatives are identified, which are difficult to rule out.

\vskip 0.5\baselineskip
\selectlanguage{francais}
\noindent{\bf R\'esum\'e}
\vskip 0.5\baselineskip
\noindent

Je propose une approche pour identifier, parmi plusieurs stratégies d'analyse
phylogénétique, celles aux résultats les plus fiables.  Cette approche se base
sur l'hypothèse que plus un résultat est reproduit à partir de données
indépendantes, plus il reflète le signal historique commun aux données
analysées.  Sous cette hypothèse, la capacité d'une stratégie d'analyse à
extraire le signal historique devrait être positivement corrélée à la cohérence
des résultats obtenus.  J'applique cette approche à une série d'analyses sur
des données empiriques, en basant la mesure de cohérence sur les distances de
Robinson-Foulds entre les arbres obtenus.  En première approximation, les
stratégies d'analyse les plus adaptées aux données produisent les résultats les
plus cohérents. Cependant, des risques de faux positifs et de faux négatifs
difficiles à écarter sont identifiés.

\vskip 0.5\baselineskip
\selectlanguage{english}
\noindent{\bf Notice}
\vskip 0.5\baselineskip
\noindent

This is the author's version of a work that was accepted for publication in
Comptes Rendus Palevol. Changes resulting from the publishing process, such as
peer review, editing, corrections, structural formatting, and other quality
control mechanisms may not be reflected in this document. Changes may have been
made to this work since it was submitted for publication.
References of the definitive version shall be provided here in an update of this author's version.

\end{abstract}


%

\end{frontmatter}


\selectlanguage{english}
\section{Introduction}
\label{sec:intro}

An important breakthrough for molecular phylogeny reconstruction has been made
with the introduction of probabilistic approaches
\citep{Felsenstein_1981,Yang_and_Rannala_1997}, directly and explicitly using
molecular evolution models.  This usually reduces the occurrences of
reconstruction artifacts, in particular in studies at large evolutionary scales
\citep[but see][]{Simmons_2012}.  In parallel with an increased availability of
data (which permits a better estimation of the parameters of complex models)
and computational power (which permits the exploration and evaluation of a
large number of possible trees), the development of probabilistic methods was
accompanied with the development of models that take into account an increasing
number of aspects of molecular evolution such as evolutionary rate
\citep{Yang_1993} or composition
\citep{Lartillot_and_Philippe_2004,Foster_2004} heterogeneities.  The accuracy
of phylogenies can also be enhanced by using character selection or recoding
techniques
\citep{Brinkmann_and_Philippe_1999,Inagaki_et_al_2004,Hassanin_et_al_2005,Goremykin_et_al_2010,Roure_and_Philippe_2011}.

However, the diversity of methods and models available makes it difficult to
decide which strategy to adopt when trying to reconstruct a phylogeny. Some
methods are available to help the phylogeneticist in this choice. For instance,
programs like jModelTest \citep{Posada_2008} use a variety of criteria to
select a model achieving a good compromise between realism and tractability.
But such readily available tools are limited to the set of models implemented
in the phylogeny programs on which they rely.
It is also common practice to compare phylogenies obtained using different
models by applying selection criteria identical to those used in \emph{a
posteriori} model selection programs, which extends these selection approaches
to arbitrary models.
Still, the model is only one aspect of the \soa{}: Data selection or recoding
techniques also need to be chosen prior to the tree construction, a program and
its specific settings have to be chosen, and support evaluation procedures can
take diverse forms. All of these aspects form the \soa{} that leads from the
raw data to an annotated tree ready for drawing phylogenetic conclusions.

An approach suitable for the choice of such integrated \soas{} could be to make
the choice \ap{}, based on their results.  A variety of analyses would be
performed, and the ones producing the most accurate results would be chosen.
This immediately raises the question as to how to evaluate the accuracy of a
phylogeny reconstruction.  Measures such as bootstrap proportions
\citep{Felsenstein_1985} or Bayesian posterior probabilities are sometimes
regarded as reliability indicators, but they must be interpreted in the limited
context of the particular dataset that has been analysed.  Other datasets may
yield different support values (or even contradictory results) and these values
do not correlate perfectly with one another \citep{Douady_et_al_2003}.
Reliability of phylogenetic relationships is arguably better estimated when
considering trees obtained from several independent datasets, and examining the
degree to which the results are reproduced across these datasets
\citep{Miyamoto_and_Fitch_1995,Chen_et_al_2003,Dettai_and_Lecointre_2004,Li_and_Lecointre_2009}.
In this context, it has been observed that the reproducibility of the results
was higher when a better modelling of the data was used
\citep{Miyamoto_et_al_1994}. This justifies a widespread practice consisting in
using more complex models and methods when the phylogeny appears more challenging to
resolve.  This also suggests that result coherence could indeed correlate
positively with accuracy.

The purpose of the present article is to report an attempt to use the \ap{}
approach for selecting strategies of phylogenetic analyses using the
reproducibility of the results as a criterion, and to discuss some potential
pitfalls of such an approach.

\section{Materials and methods}
\label{sec:matmet}

\subsection{Test data}

The \ap{} approach was tested on empirical multi-gene data assembled in the
ambit of a yet-to-be-published work on the phylogeny of Cyanobacteria and
plastids \citep{Li_et_al_inprep}. Given the large evolutionary scale, as well
as the potential existence of horizontal gene transfers, such a dataset should
provide enough reconstruction challenge so that different analytical strategies
will have different reconstruction accuracies, and show various degrees
of result coherence.

The data consists of 73 protein-coding genes from 42 Cyanobacteria, plastids or
nuclear genes of plastidial origin.  The genes were grouped in 4 sets that were
considered internally congruent and between them incongruent by the
concaterpillar program \citep{Leigh_et_al_2008}. This program performs a series
of likelihood ratio tests under a \gtrig{} model, to evaluate whether datasets
can be forced to share topologies and branch lengths or if separate trees
provide a significantly better likelihood.  Results of maximum likelihood
analyses under a \gtrig{} model should therefore provide a reference situation
where some incoherence effectively appears between the datasets. More accurate
strategies than maximum likelihood analysis under a \gtrig{} model might be
able to recover more of the history common to all datasets, for each one of
them, and therefore be characterised by a higher coherence in the results.

\subsection{Analytical strategies tested}

For each of the 4 combined datasets, a series of various \soas{} were applied.
A name is associated to each of them to facilitate reporting and discussion of
the results.

Maximum likelihood bootstrap analyses were conducted using RAxML versions 7.0.4
and 7.3.4 \citep{Stamatakis_2006} under a \gtrig{} model, with 200
pseudo-replicates of the data.  For these analyses, the original data matrices
were used, their amino-acid translations (for which a $\mathrm{CPREV} {+}
\mathrm{I} {+} \Gamma$ model was used) as well as some versions of these
matrices where diverse combinations of sites were subjected to codon-degeneracy
recodings.

A codon-degeneracy recoding is based on the replacement of codons by degenerate
versions that represent all codons coding the same amino-acid.  Nucleotides are
replaced by IUPAC ambiguity codes at codon positions where several codons for
the same amino-acid differ.

The goal of these recodings is to eliminate potentially misleading signal.
The signal considered for removal corresponds to sites involved in codon
synonymy.  Due to the relaxed selection on the nucleotide at such sites,
convergence between sequences sharing the same bias in their genome's
nucleotide composition may have happened and mislead phylogenetic
reconstruction \citep[see for instance][]{Foster_2004, Hassanin_et_al_2005,
Cox_et_al_2008, Nabholz_et_al_2011, Rota-stabelli_et_al_2013}. The most useful
of the recodings should affect mostly sites where the proportion of misleading
signal is the highest, and eliminate only a small proportion of the historical
signal.  A higher coherence of the results is expected for such recodings.

For reasons pertaining to the organization of the genetic code (that will not
be detailed here), three categories of codon positions were distinguished:
third codon positions of any amino-acid, leucine and arginine first codon
positions, serine first and second codon positions.

The maximum likelihood \soas{} were the following
(see also supplementary document available at
\url{http://dx.doi.org/10.6084/m9.figshare.732758} for the recodings):
\begin{itemize}

\item `unrecoded': The original matrix is used, without codon-degeneracy
    recoding. This strategy is used as a reference, where some degree of
    incoherence is expected between the datasets delimited by concaterpillar.

\item `degen3': The codons are replaced with degenerate versions representing
    all synonymous codons in their family, but only the third position is
    actually recoded.

\item `degen1LR': The codons are replaced with degenerate versions representing
    all synonymous codons in their family, but only at first positions of
    codons coding a leucine or an arginine.

\item `degen12S': The codons are replaced with degenerate versions representing
    all synonymous codons in their family, but only at first and second
    positions of codons coding for a serine.

\item `degen12LRS': The codons are replaced with degenerate versions
    representing all synonymous codons in their family, but only at first and
    second positions of codons coding for a leucine, an arginine or a serine.

\item `degenLR3': The codons are replaced with degenerate versions representing
    all synonymous codons in their family, but if the codon codes anything else
    than a leucine or an arginine, only the third position is actually recoded.

\item `degenS3': The codons are replaced with degenerate versions representing
    all synonymous codons in their family, but if the codon codes anything else
    than a serine, only the third position is actually recoded.

\item `degenLRS3': The codons are replaced with degenerate versions
    representing all synonymous codons in their family, but if the codon codes
    anything other than a leucine, an arginine or a serine, only the third
    position is actually recoded (since only leucine, arginine and serine are
    affected by codon synonymy at other positions than the third one, this
    actually amounts to replacing every (non-terminating) codon by the
    degenerate version representing its entire family).

\item `translated': The amino-acid translation of the matrix is used.

\end{itemize}

The maximum likelihood analysis of the original nucleotide protein-coding data
was replicated using a partitioning of the model by which parameters of the
\gtrig{} model were estimated independently for each codon position, branch
lengths being optimized jointly on the three partitions (`unrecoded\_p'
\soa{}). This also is expected to provide more accurate results.

More sophisticated molecular evolution models were used, in order to take into
account some aspects of composition heterogeneity. The $\mathrm{CAT}$ model
implemented in Phylobayes \citep{Lartillot_and_Philippe_2004} allows different
sites to evolve under different nucleotide equilibrium frequencies. The NDCH
model implemented in P4 \citep{Foster_2004} allows different branches of the
tree to evolve under different nucleotide equilibrium frequencies.

The analyses using Phylobayes version 3.2f (named `CAT') were performed as
follows: Two Markov chain Monte Carlo (MCMC) were run under a \gtrig{} model,
with the automatic stopping criterion based on the computation of convergence
statistics between two chains (`maxdiff' $>$ 0.3 and `effective size' $>$ 50,
checking every 100 cycles).

The analysis using P4 versions 0.88.r190 and 0.88.r186 (named `NDCH') was
performed as follows: Four Metropolis coupled Markov chain Monte Carlo (MCMCMC)
were run under a \gtrig{} model until the log likelihood values of the cold
chain appeared to have plateaued and the ESS sampling values were higher than
50 (preferably at least 200, but that was not always achieved).  A posterior
predictive distribution \citep{Bollback_2002} of the $\mathrm{X}^{2}$
composition heterogeneity statistic was generated using data simulated on the
Markov chain samples \citep{Foster_2004}.  This distribution was compared to
the $\mathrm{X}^{2}$ of the empirical data to evaluate the ability of the model
to account for composition heterogeneity across the tree. This comparison was
implemented as a one-tailed area probability test (with a p-value deemed
significant if smaller than 0.05). As long as the model did generate data
significantly less heterogeneous than the empirical data, additional
composition vectors were added and new analyses were performed under a
$\mathrm{GTR} {+} \mathrm{I} {+} \Gamma {+} n\mathrm{CV}$ model (where $n$ is
the total number of composition vectors), using the same procedure as above.
In practice, two composition vectors proved enough for every analysed dataset.
The results of the `NDCH' analyses therefore correspond to Bayesian analyses
under a $\mathrm{GTR} {+} \mathrm{I} {+} \Gamma {+} 2\mathrm{CV}$ model.

In order to achieve  better comparability with bootstrap analyses, for both the
`CAT' and `NDCH' \soas{}, the coherence of the results was measured using a
random selection of 200 post-burnin samples from the Markov Chain.

In addition, three other \soas{} that were expected to have a worse performance
than the `unrecoded' maximum likelihood strategy were applied: a parsimony
analysis (`pars') and two distance-based approaches, one using Jukes-Cantor
distances (`JCdist') and the other LogDet distances
\citep[`logdet',][]{Lockhart_et_al_1994}.  All three strategies were performed
using tools from the version 3.69 of the phylip package
\citep{Felsenstein_2005} with 200 bootstrap pseudo-replicates of the data.

\subsection{Coherence measure}

The coherence of the results obtained by a given \soa{} across a series of
datasets could be evaluated in various ways, and the design of a method to
accomplish such an evaluation could be a research topic in its own right.  In
the present work, the coherence was evaluated on the basis of pairwise
topological distances between trees obtained by applying the \soa{} on the
datasets. The shorter the topological distances, the more `similar' the trees,
the more coherent the results.

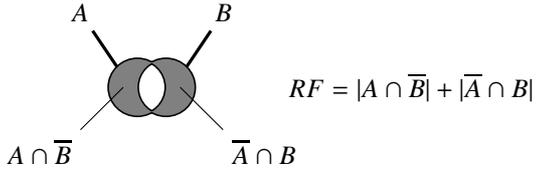
\begin{figure}[h]
\begin{tikzpicture}[scale=0.75]
    \coordinate (c1) at (6,0);
    \coordinate (c2) at (6.5,0);
    \path[name path=ligne1] (c1) -- ++ (-1,1) node[anchor=south] (A) {$A$};
    \draw[name path=cercle1] (c1) circle (.5cm);
    \draw[name intersections={of=ligne1 and cercle1},very thick] (intersection-1) -- (A);
    \draw[very thick] (c1) circle (.5cm);
    \path[name path=ligne2] (c2) -- ++ (1,1) node[anchor=south] (B) {$B$};
    \draw[name path=cercle2] (c2) circle (.5cm);
    \draw[name intersections={of=ligne2 and cercle2},very thick] (intersection-1) -- (B);
    \draw[very thick] (c2) circle (.5cm);
    \fill[even odd rule,gray] (c1) circle (.5cm) (c2) circle (.5cm);
    \path (c1) -- ++ (-0.25,0) coordinate (AnB);
    \draw (AnB) -- ++(-0.75,-0.75) node[anchor=north east] {$A \cap \overline{B}$};
    \path (c2) -- ++ (0.25,0) coordinate (nAB);
    \draw (nAB) -- ++(0.75,-0.75) node[anchor=north west] {$\overline{A} \cap B$};
    \path (c2) -- ++ (2,0) node[anchor=west] {$RF = |A \cap \overline{B}|+|\overline{A} \cap B|$};
\end{tikzpicture}
\caption{\label{RF}Robinson-Foulds symmetric difference distance ($RF$).
$A$ and $B$ represent the sets of bipartitions defined by
the branches of two trees. The grey area is the set of bipartitions that belong
to either $A$ or $B$ but not to both. This set is the symmetric difference
of $A$ and $B$. The Robinson-Foulds distance between the two trees is the
number of elements in this set.}
\end{figure}

The Robinson-Foulds symmetric difference \citep{Robinson_and_Foulds_1981} was
used as topological distance.  If we note $A$ and $B$ the sets of bipartitions
defined by the branches of two trees on a same set of leaves, the
Robinson-Foulds distance between these two trees is the number of bipartitions
present in $A$ or in $B$ but not in both (see figure \ref{RF}).

The coherence of a given \soa{} can be evaluated using the Robinson-Foulds
distances between pairs of trees obtained by applying that strategy on
different datasets sharing the same set of leaves. If the sets of leaves differ
between datasets, the trees have to be reduced to the set of common leaves.
This might raise important issues that are beyond the scope of the present
paper. Here, the four datasets delimited by concaterpillar included the same
terminal taxa.

In the present work, the \soas{} included the generation of 200
bootstrap trees or the extraction of 200 samples from a Markov chain. The
coherence was therefore evaluated using the distances between pairs of
consensus trees ($4$ consensuses of 200 trees because there are $4$ datasets,
which makes $6$ pairs) and the full distribution of the distances between pairs
of bootstrap or MCMC-sampled trees ($200^2 = 40000$ distances for a
pair of datasets, $6$ pairs of datasets).

\section{Results}
\label{sec:results}

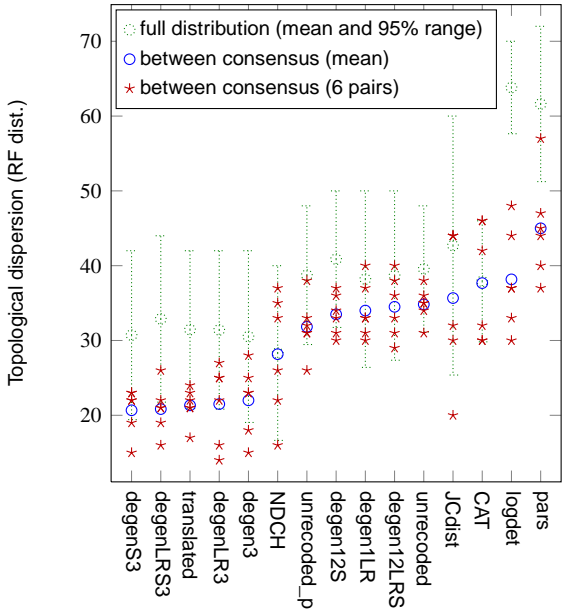
\begin{figure}[h]
\begin{tikzpicture}
\tikzstyle{every node}=[font=\footnotesize \sffamily]
\begin{axis}[height=0.33\textheight,width=75mm,enlargelimits=0.05,stack plots=false,x tick label style={rotate=-90,anchor=west},xtick={0,1,2,3,4,5,6,7,8,9,10,11,12,13,14},xticklabels={
    degenS3,%
    degenLRS3,%
    translated,%
    degenLR3,%
    degen3,%
    NDCH,%
    unrecoded\_p,%
    degen12S,%
    degen1LR,%
    degen12LRS,%
    unrecoded,%
    JCdist,%
    CAT,%
    logdet,%
    pars},ylabel=Topological dispersion (RF dist.)]
\addplot plot[only marks,mark=o,densely dotted,green,error bars/.cd,y dir=both,y explicit]
coordinates{
(0,30.71) +- (12.71,11.29)
(1,32.88) +- (12.88,11.12)
(2,31.45) +- (13.45,10.55)
(3,31.41) +- (13.41,10.59)
(4,30.52) +- (10.52,11.48)
(5,28.32) +- (12.32,11.68)
(6,38.73) +- (10.73,9.27)
(7,40.87) +- (8.87,9.13)
(8,38.20) +- (10.20,11.80)
(9,38.68) +- (8.68,11.32)
(10,39.57) +- (7.57,8.43)
(11,42.69) +- (22.69,17.31)
(12,37.91) +- (9.91,8.09)
(13,63.82) +- (13.82,6.18)
(14,61.62) +- (14.62,10.38)};

\addplot plot[only marks,mark=o,blue]
coordinates{
(0,20.67)
(1,20.83)
(2,21.33)
(3,21.50)
(4,22.00)
(5,28.17)
(6,31.83)
(7,33.50)
(8,34.00)
(9,34.50)
(10,34.83)
(11,35.67)
(12,37.67)
(13,38.17)
(14,45.00)};

\addplot plot[only marks,mark=star,red]
coordinates{
(0,22.00)
(1,21.00)
(2,24.00)
(3,25.00)
(4,28.00)
(5,22.00)
(6,38.00)
(7,33.00)
(8,37.00)
(9,40.00)
(10,36.00)
(11,44.00)
(12,42.00)
(13,44.00)
(14,45.00)};

\addplot plot[only marks,mark=star,red]
coordinates{
(0,22.00)
(1,26.00)
(2,22.00)
(3,22.00)
(4,23.00)
(5,33.00)
(6,31.00)
(7,36.00)
(8,33.00)
(9,33.00)
(10,38.00)
(11,32.00)
(12,46.00)
(13,33.00)
(14,47.00)};

\addplot plot[only marks,mark=star,red]
coordinates{
(0,19.00)
(1,16.00)
(2,21.00)
(3,27.00)
(4,23.00)
(5,37.00)
(6,31.00)
(7,34.00)
(8,40.00)
(9,31.00)
(10,34.00)
(11,44.00)
(12,32.00)
(13,30.00)
(14,37.00)};

\addplot plot[only marks,mark=star,red]
coordinates{
(0,15.00)
(1,21.00)
(2,17.00)
(3,16.00)
(4,15.00)
(5,16.00)
(6,26.00)
(7,37.00)
(8,30.00)
(9,29.00)
(10,35.00)
(11,20.00)
(12,46.00)
(13,37.00)
(14,44.00)};

\addplot plot[only marks,mark=star,red]
coordinates{
(0,23.00)
(1,22.00)
(2,23.00)
(3,25.00)
(4,18.00)
(5,26.00)
(6,32.00)
(7,30.00)
(8,31.00)
(9,38.00)
(10,31.00)
(11,30.00)
(12,30.00)
(13,48.00)
(14,57.00)};

\addplot plot[only marks,mark=star,red]
coordinates{
(0,23.00)
(1,19.00)
(2,21.00)
(3,14.00)
(4,25.00)
(5,35.00)
(6,33.00)
(7,31.00)
(8,33.00)
(9,36.00)
(10,35.00)
(11,44.00)
(12,30.00)
(13,37.00)
(14,40.00)};

\legend{full distribution (mean and 95\% range), between consensus (mean), between consensus (6 pairs)}
\end{axis}
\end{tikzpicture}
\caption{\label{results}Coherence measures for the \soas{}
tested in this study.  The vertical axis bears units of Robinson-Foulds
topological distances ($RF$), that is, numbers of bipartitions that are not
shared between a pair of trees.  The lower the value, the more similar the
trees are.  The strategies are sorted according to the mean (blue circles)
of the 6 $RF$ distances between pairs of consensus trees (red stars). Those
on the left are therefore the most coherent, and, presumably, the most
accurate. The distribution of distances between individual bootstrapped or
MCMC-sampled trees is represented by its 95\% range and its mean (in green
dots). See the text for explanations about the names of the \soas{}.}

\end{figure}

The coherence measures are reported in figure~\ref{results}.

A striking observation is that the \soas{} with the highest coherence are those
where the signal corresponding to synonymous substitutions at third codon
position has been eliminated (the names of these strategies end in `3') and
the strategy using translated data (by which an important part of the
variability corresponding to third codon positions is eliminated due to codon
synonymy). The next most coherent strategy is the `NDCH' strategy, that models
the existence of composition heterogeneities across the phylogeny.

All of these strategies appear more coherent than the reference `unrecoded'
strategy, where all signal is present in the data and a simple
composition-homogeneous \gtrig{} model is used.

This difference in performance is likely to reflect a real difference in
reconstruction accuracy. Indeed, both the suppression of the signal associated
to synonymy at third codon position and the modelling of composition
heterogeneities can reduce the risks of obtaining reconstruction artefacts
driven by convergences in genome composition biases, which we know affect
the data used here \citep{Li_et_al_inprep}.
The recoding strategies leaving third positions unaffected do not seem to lead
to a higher coherence than the `unrecoded' strategy.


Contrary to what was expected, the \soa{} where the model is partitioned by
codon position (`unrecoded\_p') does not seem to perform strikingly better than
its non-partitioned counterpart (`unrecoded').
The average $RF$ distance between bootstrap consensuses is only slightly
lower for `unrecoded\_p' than for `unrecoded'.


As expected, the \soas{} using parsimony (`pars') and distance
(`JCdist' and `logdet') seem to produce results overall less coherent than the
`unrecoded' strategy. It was however not expected that the strategy using a
site-heterogeneous model (`CAT') would also produce less coherent results than
the `unrecoded' strategy. It should be noted that the statistical significance of
this observation is not known. It may be that the `CAT' strategy reveals true
divergences between the evolutionary histories of the different dataset (see discussion).


The `NDCH' strategy used the program P4, which can produce trees with
polytomies during the Markov chain.  A tree with polytomies has less
bipartitions than a fully bifurcating tree. The lower the numbers of
bipartitions there are in a pair of trees, the lowest the upper bound on the
$RF$ distance between these trees. Indeed $RF$ is the highest when the trees
have no bipartition in common, and it is then equal to the sum of the numbers
of bipartitions found in the trees. This may bias the coherence evaluation used
in the present work.  The same problem holds for the `pars' \soa{}, because
parsimony sometimes results in a set of equally parsimonious trees, and the
(polytomous) consensus of these trees is used as the result of the analysis.

This may explain why the `NDCH' strategy had the lowest average $RF$ distances
between individual sampled trees, although the $RF$ distances between consensus
trees were not. And similarly, this may explain why the average $RF$ distance
between individual trees for the `pars' strategy was not the highest, whereas
it was the case for the distances between the bootstrap consensus trees for
this strategy.


The distances between bootstrapped trees are in average higher than the
distances between their consensuses. The bootstrapped trees for a given dataset
are by nature dispersed because they are built using different matrices that
are samples of the sites from the original alignment that possibly support
diverging topologies, but their consensus is expected to reflect the
signal present in the full alignment. If all datasets bear the same
historical structure, and provided that this structure is efficiently
revealed by the \soa{}, a better coherence is expected between
bootstrap consensuses than between the diverse bootstrapped trees.

This discrepancy in coherence between individual trees and between their
consensuses is not observed in the case of MCMC-sampled trees (`NDCH' and
`CAT').  The trees sampled from a stationary MCMC are all supposed
to be drawn from the a-posteriori probability distribution of trees given the
full data matrix.  If this distribution is centered around one highly dominant
most probable topology, the topological dispersion observed within the sample
will be low. This is often observed: be it a true characteristic of Bayesian
posterior distributions in phylogeny, or because of prevalent inefficiencies in
MCMC mixing \citep{Lakner_et_al_2008}, consensus trees obtained from MCMC
samples often have high node posterior probabilities \citep{Douady_et_al_2003}.

\section{Discussion}
\label{sec:discus}

\begin{figure}[h!]
\pgfoonew \tree=new treeplet(0.5)
\pgfoonew \uree=new treeplet(0.33)
\pgfoonew \vree=new treeplet(0.33)

\pgfoonew \targ=new target(0.5)
\pgfoonew \uarg=new target(0.33)
\pgfoonew \varg=new target(0.33)

\pgfmathsetseed{12343}

\begin{tikzpicture}[scale=0.6]
    \node[inner sep=0.5cm] (t) at (0,0) {};
    \tree.draw(t)
    \path (t) -- ++(0,-2) node (letter) {a};

    \path (t) -- ++(4,0) node (start) {datasets};
    \draw[->] (t) -- (start);
    
    \path (t) -- ++(10,0) coordinate (c);
    \targ.draw(c)
    \path (c) -- ++(0,2) node (results) {results};
    \path (start) -- (results) coordinate[midway] (method);
    \path (method) -- ++(0,1) node[align=center] {{\color{green}`good' strategy}\\{\color{red}`bad' strategy}};
    
    \path (c) -- ++(rand*0.1,rand*0.2) node[inner sep=1pt,circle,fill,green] (good1) {};
    \path (c) -- ++(rand*0.1,rand*0.2) node[inner sep=1pt,circle,fill,green] (good2) {};
    \path (c) -- ++(rand*0.1,rand*0.2) node[inner sep=1pt,circle,fill,green] (good3) {};
    \path (c) -- ++(rand*0.1,rand*0.2) node[inner sep=1pt,circle,fill,green] (good4) {};
    \draw[->,thin,green] (start) -- (good1);
    \draw[->,thin,green] (start) -- (good2);
    \draw[->,thin,green] (start) -- (good3);
    \draw[->,thin,green] (start) -- (good4);

    \path (start) -- (results) coordinate[midway] (method);
    
    \path (c) -- ++(rand*0.5,rand*1) node[inner sep=1pt,circle,fill,red] (bad1) {};
    \path (c) -- ++(rand*0.5,rand*1) node[inner sep=1pt,circle,fill,red] (bad2) {};
    \path (c) -- ++(rand*0.5,rand*1) node[inner sep=1pt,circle,fill,red] (bad3) {};
    \path (c) -- ++(rand*0.5,rand*1) node[inner sep=1pt,circle,fill,red] (bad4) {};
    \draw[->,thin,red] (start) -- (bad1);
    \draw[->,thin,red] (start) -- (bad2);
    \draw[->,thin,red] (start) -- (bad3);
    \draw[->,thin,red] (start) -- (bad4);
\end{tikzpicture}

\begin{tikzpicture}[scale=0.6]
    
    \node[inner sep=0.5cm] (t) at (0,0) {};
    \path (t) -- ++(0,0.75) node[inner sep=0.33cm] (u) {};
    \path (t) -- ++(0,-2) node (letter) {b};
    \uree.draw(u)
    \path (t) -- ++(0,-0.75) node[inner sep=0.33cm] (v) {};
    \begin{scope}[xscale=-1]
    \vree.draw(v)
    \end{scope}
    \path (t) -- ++(4,0) node (start) {};
    \path (u) -- ++(4,0) node (ustart) {datasets};
    \draw[->] (u) -- (ustart);
    \path (v) -- ++(4,0) node (vstart) {datasets};
    \draw[->] (v) -- (vstart);
    \path (t) -- ++(10,0) coordinate (c);
    \path (u) -- ++(10,0) coordinate (d);
    \uarg.draw(d)
    \path (v) -- ++(10,0) coordinate (e);
    \varg.draw(e)
    \path (c) -- ++(0,2) node (results) {results};
    \path (start) -- (results) coordinate[midway] (method);
    \path (method) -- ++(0,1) node[align=center] {{\color{green}`good' strategy}\\{\color{red}`bad' strategy}};

    \path (d) -- ++(rand*0.1,rand*0.2) node[inner sep=1pt,circle,fill,green] (ugood1) {};
    \path (d) -- ++(rand*0.1,rand*0.2) node[inner sep=1pt,circle,fill,green] (ugood2) {};
    \path (e) -- ++(rand*0.1,rand*0.2) node[inner sep=1pt,circle,fill,green] (vgood1) {};
    \path (e) -- ++(rand*0.1,rand*0.2) node[inner sep=1pt,circle,fill,green] (vgood2) {};
    \draw[->,thin,green] (ustart) -- (ugood1);
    \draw[->,thin,green] (ustart) -- (ugood2);
    \draw[->,thin,green] (vstart) -- (vgood1);
    \draw[->,thin,green] (vstart) -- (vgood2);

    \path (d) -- ++(rand*0.5,rand*1) node[inner sep=1pt,circle,fill,red] (ubad1) {};
    \path (d) -- ++(rand*0.5,rand*1) node[inner sep=1pt,circle,fill,red] (ubad2) {};
    \path (e) -- ++(rand*0.5,rand*1) node[inner sep=1pt,circle,fill,red] (vbad1) {};
    \path (e) -- ++(rand*0.5,rand*1) node[inner sep=1pt,circle,fill,red] (vbad2) {};
    \draw[->,thin,red] (ustart) -- (ubad1);
    \draw[->,thin,red] (ustart) -- (ubad2);
    \draw[->,thin,red] (vstart) -- (vbad1);
    \draw[->,thin,red] (vstart) -- (vbad2);
\end{tikzpicture}

\begin{tikzpicture}[scale=0.6]
    \node[inner sep=0.5cm] (t) at (0,0) {};
    \tree.draw(t)
    \path (t) -- ++(0,-2) node (letter) {c};

    \path (t) -- ++(4,0) node (start) {datasets};
    \draw[->] (t) -- (start);
    
    \path (t) -- ++(10,0) coordinate (c);
    \targ.draw(c)
    \path (c) -- ++(0,2) node (results) {results};
    \path (start) -- (results) coordinate[midway] (method);
    \path (method) -- ++(0,1) node[red] {`bad' strategy};
    
    \path (c) -- ++(rand*0.5,rand*1) coordinate (f);

    \path (f) -- ++(rand*0.1,rand*0.2) node[inner sep=1pt,circle,fill,red] (biased1) {};
    \path (f) -- ++(rand*0.1,rand*0.2) node[inner sep=1pt,circle,fill,red] (biased2) {};
    \path (f) -- ++(rand*0.1,rand*0.2) node[inner sep=1pt,circle,fill,red] (biased3) {};
    \path (f) -- ++(rand*0.1,rand*0.2) node[inner sep=1pt,circle,fill,red] (biased4) {};
    \draw[->,thin,red] (start) -- (biased1);
    \draw[->,thin,red] (start) -- (biased2);
    \draw[->,thin,red] (start) -- (biased3);
    \draw[->,thin,red] (start) -- (biased4);
\end{tikzpicture}
\caption{\label{targets}Possible cases for the coherence of results.
The symbolic trees on the left represent evolutionary histories.
The targets on the right represent the space of results.
The datasets in the middle are the product (black arrow) of an evolutionary
history, and are used to infer a result (dot on a target whose centre represent
the true relationships) according to a given \soa{} (green or red arrow).
The more compact a set of dots, the more coherent the results it represents.
a: The datasets are the products of a same evolutionary history and there is no
systematic bias repeatedly affecting the reconstruction.  A `good' \soa{}
will generate results close to the true historical relationships.
A `bad' \soa{} will produce more dispersed results.
b: The datasets are the products of different evolutionary histories.  A `good'
\soa{} will produce results close to the historical
relationships having produced the dataset on which the inference is based.
The results will therefore be situated in different areas of the
result space. The distribution of the results obtained by a `good' \soa{}
will be less easy to distinguish from the results obtained by a `bad' \soa{} than
when the datasets are the products of a same history.
c: The datasets are the products of a same evolutionary history and the \soa{}
is sensitive to a systematic inference error. The results will tend to
concentrate around a zone of the result space corresponding to an artefactual
reconstruction. This artefactual reconstruction will therefore correspond to a
coherent set of results, and may be mistaken for the true evolutionary history.}

\end{figure}
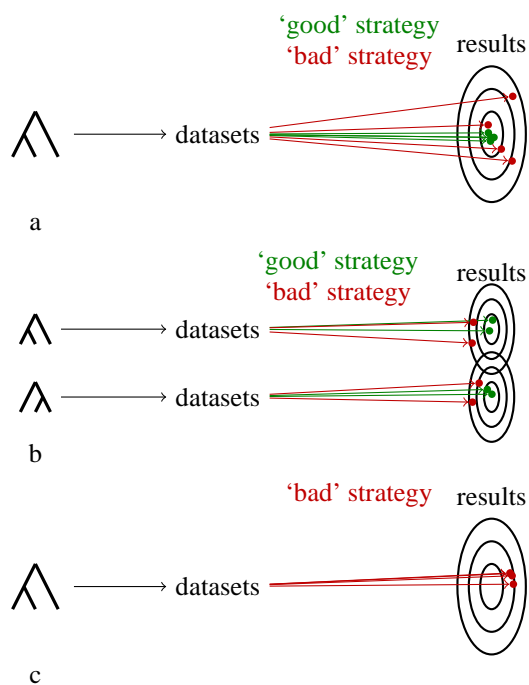

The most accurate strategies should extract more historical signal than the
others.  Consequently, independent datasets issued from a same evolutionary
history should produce more similar trees when using these strategies than when
using less accurate \soas{} (figure~\ref{targets}a). This seems to be the case
to some extent with our data, because the \soas{} that are designed to counter
the misleading effects of convergence in composition bias (`degen*3',
`translated', and `NDCH') tend to produce more coherent topologies than the
other strategies. If the accuracy improvement brought by these strategies is
important enough, the hypothesis can be made that their use in concaterpillar
(instead of \gtrig{} maximum likelihood analyses on unrecoded data) could yield
a smaller number of combined datasets, due to a higher compatibility between
the phylogenies supported by the individual protein-coding genes (and their
combinations) under strategies more apt to overcome reconstruction artefacts.

The coherence-based \ap{} approach will probably not perform well when the
different datasets were generated on different histories.  A good \soa{} should
produce coherent results only within a set of datasets sharing the same
history. Otherwise, result dispersion is expected. It is difficult to tell to
what extent such dispersion will be strong relative to dispersion due to the
use of inaccurate strategies.  Such strategies might then be difficult to
distinguish from more accurate ones (figure~\ref{targets}b). The surprisingly
high relative incoherence obtained using the `CAT' strategy may be due to the
existence of conflicts between datasets that can only be revealed when
heterogeneity of composition across sites is taken into account. Such conflicts
are not unexpected, due to the bacterial nature of the taxa used for the
present tests. The type of pitfall described in figure~\ref{targets}b may be
less a problem with organisms not so prone to horizontal transfers of genetic
material.

But even in cases where datasets all result from the same history, a potential
pitfall exists that depends in the way the less accurate \soas{} are
inaccurate.  If they are inaccurate in a systematically biased way, then the
trees may be similar due to shared wrongly inferred relationships
(figure~\ref{targets}c). Including `control' \soas{} chosen for their known
sensitivity to some specific systematically biased artefacts could help the
detection of false positives due to these biases. If a control \soa{} happens
to produce more coherent results than other strategies a priory more likely to
be accurate, this is a hint that the reconstruction artefact can affect some
analyses, and that the coherence criterion should be considered with caution.

The coherence measure used here is based on $RF$ distances, which makes it
possibly biased because \soas{} producing less resolved trees might never look
as incoherent as strategies producing fully bifurcating trees.  However, it is
not sure that this is an undesired property. Indeed, one might want to favour
\soas{} that are conservative and avoid displaying relationships not well
enough supported.

\section{Conclusion}

The coherence-based \ap{} approach tested here seems to behave partially as
expected. Some improvements can probably be made in the way coherence is
measured, but the potential pitfalls to which it might be sensitive seem
difficult to rule out.  The presented approach may not be suitable as a good
selection tool in practice. Different \soas{} may be equally
coherent but for different (bad or good) reasons. It may therefore be advisable
to use such a method as an exploratory tool rather than as a decision tool.

\section*{Acknowledgements}
\label{sec:ack}

Thanks to João Sollari Lopes and Cymon J. Cox for assembling and providing the
datasets that were used in this study.
Thanks to Michel Laurin and the SFS for giving me the opportunity to present
these ideas at the `Journées d'automne 2012 de la SFS'.
Thanks to the editor an reviewers for their comments and advice that hopefully
helped me improving the present paper.

This work was supported by a Fundação para a Ciência e a Tecnologia (FCT,
Portugal) grant to Cymon J. Cox, Centro de Ciências do Mar (CCMAR) - CIMAR-Lab.
Assoc., (PTDC/BIA-BCM/099565/2008).

\bibliographystyle{model2-names}
\bibliography{journ,biblio}







\end{document}